\documentclass[conference]{IEEEtran}
\usepackage{adjustbox}
\usepackage{graphicx}
\usepackage{array}
\usepackage[skip=0pt]{subcaption}
\usepackage[labelformat=simple]{subcaption}

\usepackage{amsmath}
\usepackage{amssymb}
\usepackage{algorithm}
\usepackage{algpseudocode}
\usepackage{url}
\usepackage{tikz}
\usepackage{balance}
\usepackage{multirow}
\usepackage{comment}
\usepackage{textcomp}
\usepackage{euscript}
\usepackage{makecell}
\usepackage[symbol*]{footmisc}
\usepackage[skip=0pt]{caption}
\usepackage{leading}
\leading{12pt}
\usepackage{etoolbox}

\newcommand{\isu}{$^\dagger$}
\newcommand{\eurecom}{$^\S$}
\newcommand{\osa}{$^\P$}

\usepackage{fancyhdr}
\fancypagestyle{firstpage}{  
    \fancyhf{}  
    \fancyhead[L]{\footnotesize This paper has been accepted for publication on the 11th IEEE International Conference on Network Softwarization (IEEE NetSoft 2025). This is the authors' accepted version of the article. Once it is published in IEEE under their copyright, this version may no longer be available. }  
}

\begin{document}

\title{AraRACH: Enhancing NextG Random Access Reliability in Programmable Wireless Living Labs}
\vspace{-5mm}

\author{ 
\IEEEauthorblockN{Joshua Ofori Boateng\isu, Tianyi Zhang\isu, Guoying Zu\isu, Taimoor Ul Islam\isu, Sarath Babu\isu, Florian Kaltenberger\eurecom,\\ Robert Schmidt\osa, Hongwei Zhang\isu, and Daji Qiao\isu}
\IEEEauthorblockA{\isu Iowa State University, Ames, IA, USA}

\IEEEauthorblockA{\eurecom EURECOM, Sophia Antipolis, France; \eurecom Northeastern University, Boston, MA, USA}
\IEEEauthorblockA{\osa OpenAirInterface Software Alliance, Sophia Antipolis, France}
\IEEEauthorblockA{\isu \{jboateng, tianyiz, gyzu, tislam, sarath4, hongwei, daji\}@iastate.edu,\\ \eurecom florian.kaltenberger@eurecom.fr, \osa robert.schmidt@openairinterface.org}
}
\maketitle{}
\thispagestyle{firstpage}

\begin{abstract}

The rapid evolution of wireless technologies has intensified interest in open and fully programmable radio access networks for whole-stack research, innovation, 
and evaluation of emerging solutions. Large-scale wireless living labs, such as ARA, equipped with real-world infrastructure play a vital role in this evolution by enabling researchers to prototype and evaluate advanced algorithms for next-generation wireless systems in outdoor and over-the-air environments benefiting from real-world fidelity and end-to-end programmability. However, at the core of this innovation is the performance in terms of coverage and reliability of these wireless living labs. For instance, interfacing power amplifiers and low noise amplifiers with software-defined radios~(SDRs) for experimenting outdoors introduces issues in random access procedure---a process crucial in establishing connectivity between user equipment ~(UE) and the core network in 5G and 6G systems. Therefore, to ensure seamless
connectivity and reliable communications in open-source 5G software stacks such as OpenAirInterface~(OAI), 
we propose a slot-based approach to the 5G random access procedure leveraging full downlink~(DL) and uplink~(UL) slots instead of using special or mixed slots. We highlight how 
this approach 
achieves reliable 5G connectivity over 1\ mile---the longest 
communication range that has been achieved so far in real-world settings using open-source 5G software stacks and 
the Universal Software Radio Peripheral~(USRP) SDRs. 
We also demonstrate that, in a highly obstructed environment such as an industrial setting, we can increase the probability of a successful random access procedure to 90\%--100\% when we use at least 9 OFDM symbols to transmit \texttt{msg2} and \texttt{msg3}. 

\end{abstract}

\begin{IEEEkeywords}
ARA Wireless Living Lab, software-defined radio, end-to-end programmability, open-source, 5G, OpenAirInterface, random access procedure 
\end{IEEEkeywords}

\section{Introduction}
\label{sec:introduction}

Next-generation cellular networks, including 5G and 6G, 
aim to deliver unparalleled capacity, ultra-reliable 
connectivity, and exceptionally low latency to support a wide range of applications. Realizing these services requires a collaborative ecosystem, one driven by innovation, open-source contributions, and active research. Leading open-source radio access network~(RAN) projects such as OpenAirInterface~(OAI)~\cite{OpenAirI65:online} and srsRAN~\cite{srsRANPr14:online} play a crucial role in democratizing 5G network development, research and prototyping, providing accessible platforms for comprehensive whole-stack research and experimentation. By utilizing OAI and srsRAN, researchers can customize software stacks to explore and test various aspects of 5G, including novel algorithms, protocols, and network configurations—capabilities often limited in proprietary solutions.

In the same vein, wireless living labs with 
large-scale, fully programmable real-world 
testbeds play an essential role in shaping the future of wireless networks, acting as practical, high-fidelity 
environments where new technologies can be tested, refined, and validated. One of the core challenges of these open wireless testbeds is ensuring robust and reliable experimentation performance, particularly in the context of 5G and 6G networks. Key challenges to 
developing 
large-scale wireless living labs 
include the complex impacts that real-world systems and environmental factors such as varying weather conditions, diverse terrains, and different interference levels have on wireless channel and communication behaviors. These impacts have not been well addressed in existing open-source 5G/6G platforms, thus preventing the research and innovation communities from studying 5G/6G systems in open, real-world experimental testbeds. 

To fully realize the potential of these wireless living labs, 
there is a critical need to support open-source 5G and 6G stacks designed for reliable, end-to-end, and whole-stack prototyping. These testbeds form 
essential platforms for experimental research and prototyping. 
They enable researchers to test, validate, and assess the performance of new technologies~\cite{10.1145/2980159.2980163}. Therefore, it is necessary for these systems to leverage open-source user equipment such as OAI UE and srsUE rather than relying on commercial off-the-shelf (COTS) UEs. Open-source UEs, together with open-source gNodeBs~(gNBs), offer unmatched flexibility, enabling researchers to customize, modify, and experiment with every layer of the protocol stack. This capability is invaluable for exploring new technologies, refining network designs, and validating cutting-edge algorithms.
Leveraging SDRs such as the Universal Software Radio Peripherals~(USRPs) as radio frontend units for open-source RAN platforms is essential for pushing the boundaries of 6G research and development towards open and fully programmable RANs---a goal that can be realized if open-source platforms can achieve high levels of reliability comparable to existing commercial RANs. Such high reliability is particularly critical for maintaining consistent performance under real-world conditions, such as varying interference levels, diverse environmental scenarios, and high mobility.
By matching the performance of commercial RANs, open-source and software-defined platforms can bridge the gap between research and deployment, foster innovation, and empower the research community to tackle 6G challenges head-on, paving the way for groundbreaking advancements in cellular systems.

Achieving high reliability in seamless connectivity and wide coverage in open-source RAN software stacks (e.g., 
OAI nrUE and OAI gNB) with USRPs and RF frontends (e.g., 
power amplifiers~(PAs) and low noise amplifiers~(LNAs)) can be daunting, as we have observed in 
field-deployed, fully-programmable wireless living labs 
such as ARA~\cite{ARA:Design-Impl}. 
One major problem when deploying open-source RANs with USRPs, and power amplifiers at both UE and gNB on large-scale testbeds is the random access procedure---a crucial step necessary for the UEs to get 
attached to the network.

In this paper, we first present the random access procedure problem in open-source 5G/6G 
protocol stacks such as OAI and study how the use of special slots to schedule \texttt{msg2} and \texttt{msg3} impacts the success of the 5G random access procedure
in real-world settings, especially leveraging the fully programmable ARA wireless living lab. 
Secondly, we present a solution to the problem as the first step toward enabling seamless whole-stack open-source nextG and open RAN research, testing, and experimentation on large-scale and field-deployed wireless testbeds using USRPs.
The key contributions in this work are as follows:
\begin{enumerate}
    \item To interface power amplifiers and low-noise amplifiers with SDRs, we present in detail how open-source 5G and nextG software stacks such as OAI leverage USRP hardware driver APIs to drive GPIO control signals for time division duplex~(TDD) RF frontend TX/RX switching.
    \item To the best of our knowledge, this is the first work 
    to present and solve the random access procedure problem resulting from utilizing the special time slots in 
    5G random access on a fully programmable next-generation wireless living lab. 
    \item To ensure reliable connectivity between field-deployed open-source UEs and gNBs, we present the approach 
    of utilizing full DL and UL time slots for 5G random access procedure and analyze its benefits as opposed to the use of special time slots. Besides utilizing full DL and UL slots, we show that the start symbols and length values used to schedule \texttt{msg2} and \texttt{msg3} have an impact on the success of the 5G random access procedure in a real-world setting.
    \item By leveraging the ARA wireless living lab, we demonstrate how 
    time-varying wireless channels affect the detection and decoding of \texttt{msg2} and corresponding transmission of \texttt{msg3} for different start symbol and length value combinations. 
\end{enumerate}

\section{Related Work}
\label{sec:related_work}

Due to their crucial role in testing, experimentation, and prototyping of wireless systems, numerous wireless testbeds have been deployed worldwide with specific focuses. For instance, Niigata University in Japan implemented a wireless mesh network testbed in rural mountainous areas~\cite{takahashi2007wireless} while the Converged Infrastructure for Emerging Regions~(CIER) developed a wireless mesh network testbed in Finland, emphasizing energy efficiency and cost-effectiveness~\cite{CIER}. A 5G platform at the University of Bristol~\cite{harris2015BIO}, as part of the Bristol Is Open~(BIO) city testbed, was designed specifically for smart city applications.  The 5G RuralFirst testbed~\cite{5GRuralF67online}, situated in a rural area of the UK, facilitates experimentation across a wide range of use cases, including dynamic spectrum sharing, broadcasting, agriculture, and industrial IoT. While the above-mentioned wireless testbeds  undoubtedly accelerated advancements in wireless networking technologies, they lack support for open-source  fully programmable end-to-end wireless systems.

Testbeds equipped with SDRs offer researchers with opportunities to conduct end-to-end 5G or Open RAN experiments. Examples include Colosseum~\cite{9677430}, the world’s largest wireless network emulator, featuring 256 software-defined radios and developed by Northeastern University in Boston, USA. Arena~\cite{BERTIZZOLO2020107436}, employing 24 SDRs and 64 antennas mounted on the ceiling of a 2,240-square-foot office space, is primarily designed for spectrum research. The Drexel Grid SDR Testbed~\cite{8824901} includes several dozen N210/X310 NI SDRs deployed in a ceiling-mounted network, while Patras 5G~\cite{tranoris2019patras} offers a private 5G network for testing and experimentation, incorporating an open-source core and open-source UEs and g/eNB, and SDRs. However,  such testbeds are deployed indoors without power amplifiers in their RF frontends and, therefore, do not account for the real-world, time-varying outdoor wireless channels that are critical for field prototyping of future wireless technologies.

Unlike indoor testbeds,  some provide field-deployed SDRs. For example, NITOS~\cite{pechlivanidou2014nitos}, one of Europe’s largest single-site open experimental facilities, supports Wi-Fi, WiMAX, and 5G experimentation and includes 10 SDRs each in both its indoor and outdoor setups. POWDER~\cite{breen2020powder}, with a strong focus on SDR, enables software-programmable experimentation on 5G and beyond, massive MIMO, ORAN, spectrum sharing, CBRS, and RF monitoring. AERPAW~\cite{marojevic2020aerpaw}, an aerial experimentation platform for wireless research, supports communications via SDRs on fixed base stations and drones. COSMOS~\cite{raychaudhuri2020cosmos}, a city-scale advanced wireless testbed, spans one square mile in New York City. However, these testbeds do not employ programmable and TDD-compliant power amplifiers and low-noise amplifiers at both gNB and UE, which are necessary for reliable end-to-end 5G experiments with USRPs. Moreover, while some utilize commercial open radio units~(O-RUs), these do not allow fully programmable 5G experimentation from UE to gNB.


AraSDR~\cite{arasdr-2024} presented the design and implementation of a fully programmable 5G network with USRPs, 
along with programmable power amplifiers~(PAs) and low noise amplifiers (LNAs) to enhance signal power and range over several meters in the ARA wireless living lab. However, the deployment introduces challenges in the random access (RA) procedure---a fundamental process for establishing initial connectivity between user equipment~(UE) and the core network. None of the aforementioned works have addressed the problem of random access procedures in open-source 5G and nextG protocol stacks, particularly in large-scale field deployments using only Software-Defined Radios~(SDRs) with PAs and LNAs.


\section{OpenAirInterface Integration with RF Frontend and Random Access Implementation}
\label{sec:oai_signal_transmission}

\subsection{OAI Signal Transmission and Reception with RF Frontend} 

The transmission and reception of 5G signals with USRPs are relatively easier indoors than outdoors. In indoor environments without any amplifier, the 5G~software stack directly transmits OFDM signals from the transmit antenna and receives the signals using the receive antenna. However, in the case of real-world outdoor deployments, the signal transmission procedure becomes relatively complex with the involvement of RF front-end, such as power amplifiers and low noise amplifiers, in the loop.
While deploying an end-to-end fully programmable 5G network that leverages open-source 5G protocol stacks, such as OAI and SDRs, along with amplifiers, a separate communication session is established between the SDR and the amplifier for signal transmission and reception. The communication session is created specifically for TDD scenarios where the same frequency channel is used for transmission and reception.
To facilitate this dedicated communication, the USRPs are equipped with onboard General Purpose Input/Output~(GPIO) pins that send a control signal to the GPIO pins on the amplifier to switch to transmission or reception mode depending on the TDD configuration.

\begin{figure}[t!]
    \centering
    \includegraphics[width=0.75\columnwidth]{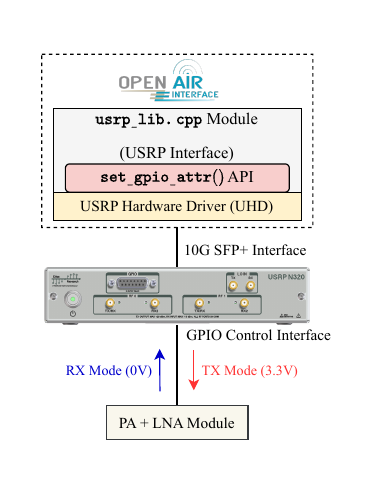}\vspace{-7mm}
    \caption{RF Frontend Control using GPIO Signal in OAI
    }
    \label{fig:gpio}
\end{figure}

Open-source 5G protocol stacks such as OAI leverage the USRP Hardware Driver~(UHD) API functions to send control signals to the RF~front-ends to toggle between transmission and reception. The GPIO pins on the USRPs are controlled by the FPGA's Automatic Transmit/Receive~(ATR) function. Fig.~\ref{fig:gpio} shows how the OAI software stack controls the RF front end in TDD mode. During initialization, OAI leverages the USRP interface module to set the USRP GPIO pins as outputs to drive the RF amplifier into either transmission or reception mode. The USRP interface, specifically \texttt{usrp\_lib.cpp}, is a software module that sits between the PHY layer and the SDR RF front-end. It implements functions that rely on UHD APIs to configure the USRP (GPIO, channel, bandwidth, gains, etc.) and perform RF I/O operations (i.e., send and receive time-domain I/Q samples)~\cite{9492647}. The USRP interface module makes the \texttt{set\_gpio\_attr()} API function call to UHD to configure the Data Direction Register~(DDR) to set specific GPIO pins as outputs using the \texttt{DDR} attribute. This allows the USRP to drive these pins during transmission or reception. To instruct the USRP to automatically manage the selected pins, OAI calls \texttt{set\_gpio\_attr()} API with \texttt{CTRL} attribute. Finally, the \texttt{ATT\_XX} is used to drive the GPIO pins when the USRP is in full-duplex mode (i.e., transmitting and reception).

The OAI CU/DU protocol stack is always in reception mode by default. A high~3.3V control signal is sent from the USRP GPIO pins to switch the amplifier to transmission mode (i.e., PA activated). On the other hand, a low 0V signal is sent to switch the amplifier to reception mode (i.e., LNA activated).

\subsection{Random Access Procedure in OpenAirInterface5g}

\begin{figure}[h!]
\hspace{-7mm}
\includegraphics[width=\columnwidth]{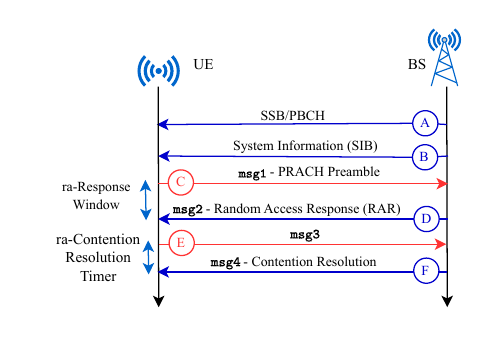}\vspace{-5mm}
    \caption{Contention-Based Random Access Procedure in 5G}
    \label{fig:ra}
\end{figure}

Random access procedure plays a fundamental role within the 5G NR protocol stack, providing the process 
needed for a UE to initiate communications with the gNodeB, synchronize timing, 
and effectively manage access contention. Generally, the 5G random access procedure could be either contention-based or contention-free. The contention-based procedure allows the UE to select a random access preamble from a pool of preambles shared with other UEs. That is, multiple UEs can select the same preamble, leading to contention. In the contention-free procedure, the base station allocates a dedicated random access preamble to a UE. OpenAirInterface5g~(OAI) specifically implements contention-based random access~(CBRA) as illustrated in Fig.~\ref{fig:ra}. In what follows, we delve into the implementation of random access in OAI and 5G NR in general.

Downlink synchronization is an integral part of the random access procedure implementation. It is explained in Steps~A and B as follows: The gNB broadcasts a synchronization signal block~(SSB) in Step~A. The SSB contains synchronization signals and the Physical Broadcast Channel~(PBCH) carrying the necessary information required by the UE to access the 5G NR cell. In Step~B, the gNB transmits the System Information Block~(SIB) to the UE, which includes the necessary information and parameters for the initial attach. 
SIB includes transmission parameters for the Physical Random Access Channel~(PRACH) configuration consisting of the PRACH preamble format, as well as time and frequency resources. PRACH denotes the physical channel carrying the UE's preambles to the base station. 
OAI 5G stack utilizes CBRA, which is a 4-step process explained in Steps~C to~F below. The CBRA approach for random access begins when the UE selects the random access preamble randomly from a pool of predefined preambles. In general, every PRACH occasion offers a maximum of 64 preambles, numbered from 0 to 63. This pool contains contention-based, contention-free preambles as well as preambles reserved for on-demand service information requests~\cite{johnson20195g}. The \texttt{ssb-perRACH-OccasionAndCB-PreamblesPerSSB} parameter broadcasted by the gNB, as part of the SIB, defines the number of synchronization signals~(SS)/PBCH beams share the same PRACH occasion and the number of preambles assigned to each SS/PBCH beam for contention-based random access. For instance, OAI, by default, uses a single PBCH beam per PRACH occasion and 60 preambles for contention-based random access. This means that the remaining set of 4 preambles contains both contention-free and reserved preambles. 
The UE, after randomly selecting a preamble from the pool, transmits it together with the sequence number for the preamble on the PRACH in Step~C. After \texttt{msg1} reception, the gNB responds with \texttt{msg2}/Random Access Response~(RAR) transmission within a period, specified by ra-ResponseWindow, in Step~D. It is worth noting that multiple UEs can select the same preamble. If this happens, those UEs decode the same content from the \texttt{msg2} sent by the gNB. Each UE, after receiving the RAR, decodes it and sends \texttt{msg3} in Step~E on the same resource blocks and symbols after a period of $k_2$ plus 
$\delta$, depending on the numerology. The gNB receives and decodes a single \texttt{msg3} from only one UE. Steps C, D, and E are necessary for UL synchronization and scheduling between the gNB and UE. The final step of the CBRA procedure is the contention resolution step, where the gNB transmits \texttt{msg4}~(Contention Resolution) to the UE whose \texttt{msg3} was successfully decoded in Step~F. After decoding \texttt{msg4}, the successful UE discards the contention resolution timer, and the random access procedure is considered successful. However, the unsuccessful UEs restart the procedure with another preamble transmission.



\subsection{TDD UL/DL Common Configuration}
\label{subsec:tdd_ul_dl_config}
The TDD UL/DL common configuration defines the uplink and downlink configuration for a TDD system. In other words, we must clearly define when and within which slot 
to expect a transmission or reception. The \texttt{TDD-UL-DL ConfigurationCommon} parameter in OAI 
is crucial for the message exchanges between gNB and UE during the RACH procedure. The parameters used to specify this configuration are the period, the number of slots in a radio frame, and the number of symbols in a slot. In 5G new radio, downlink and uplink transmissions are organized into radio frames with  duration 10\,ms, each consisting of ten subframes of 1\,ms~\cite{5GNR_numerology}. Within each subframe are slots whose length or duration depends on the subcarrier spacing~(scs). For instance, with 30\,kHz scs shown in TABLE ~\ref{tab: tdd_params}, the length of the slot is 0.5\,ms, which translates into two slots per subframe. Depending on the cyclic prefix~(CP), the number of OFDM symbols can either 14 or 12 (i.e., 14~OFDM symbols for normal CP and 12~for extended CP). In OAI, there exist 14~OFDM symbols in each slot.

\vspace{-5mm}
\begin{figure}[h!]
\hspace{2mm}
    \includegraphics[width=\columnwidth]{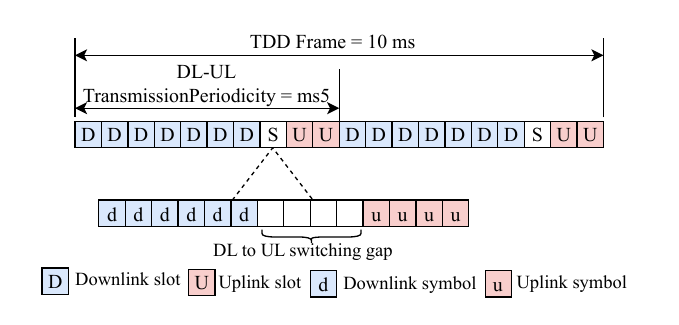}
    \caption{TDD UL/DL Common Configuration}
    \label{fig:tddconfig}
\end{figure}
 In Fig.~\ref{fig:tddconfig}, the DL-UL-TransmisionPeriodicity represents the period of the DL-UL pattern comprising of full DL slots followed by a special slot, with both DL and UL symbols. The special slot is followed by full UL slots.
 Depending on the periodicity, there may exist a guard period between the DL and UL symbols of the special slot. The guard period is crucial 
 for DL/UL switching in TDD.
 OAI 5G code utilizes \texttt{ms5} periodicity, with the default \texttt{DDDDDDDSUU} DL-UL pattern, which translates into 7 DL slots, 1~special/mixed slot, and 2~UL slots as shown in TABLE~\ref{tab: tdd_params}.

\begin{table}[h!]
\centering
\caption{Default OAI TDD Parameters}
\renewcommand{\arraystretch}{1.2} 
\begin{adjustbox}{max width=1.0\columnwidth}
\begin{tabular}{| l | c |}
\hline
\textbf{Parameter}          & \textbf{Description} \\ \hline\hline
TDD configuration  & DDDDDDDSUU  \\\hline
Subcarrier spacing & 30 kHz       \\\hline
DL symbols         & 6           \\\hline
UL symbols         & 4           \\\hline
DL-UL-Periodicity  & ms5         \\ \hline
\end{tabular}
\end{adjustbox}
\label{tab: tdd_params}
\end{table}

\section{Scheduling \texttt{msg2} and \texttt{msg3} in Special Slots: Challenges}
\label{sec:special_slot_challenges}


The random access procedure
begins with the gNodeB~(gNB) broadcasting the first Radio Resource Control~(RRC) message, also known as System Information Block~(SIB). 
The SIB 
message contains and provides RACH-related information, such as RACH ConfigCommon, to the UE to begin transmitting random access preambles on the PRACH.
Specifically, OAI schedules RA preambles \texttt{msg1} to be transmitted in the uplink slot,  \texttt{slot 19}. Following \texttt{msg1} reception with a particular preamble index, the gNB begins scheduling \texttt{msg2} Downlink Control Information~(DCI). The DCI contains the information necessary to allocate physical resources for downlink and uplink data transmissions, as well as to adjust the uplink power for power control~\cite{5GShareT66:online}. By default, the OAI gNB schedules \texttt{msg2} DCI, specifically DCI format \texttt{1\_0}, to be transmitted in a special slot \texttt{slot 7} on the Physical Downlink Control Channel~(PDCCH) which is a key component of the physical layer in 5G NR that carries DCI from gNB to UE. 
The special slot is configured with 6 downlink, 4 silent, and 4 uplink OFDM symbols as illustrated in Fig.~\ref{fig:tddconfig}.

\subsection{Monitoring and Detection of DCI Format \texttt{1\_0} without RF Frontend}
Depending on the 5G NR deployment architecture, the USRP-based 5G NR UE may or may not detect the DCI format \texttt{1\_0}. DCI format \texttt{1\_0} is a specific type of downlink control information message that the network sends to a UE to indicate the allocation of physical downlink shared channel~(PDSCH) resources. The PDSCH contains the network's response to the UE's RACH preamble, including information such as timing advance and allocated uplink resources. To accurately identify the intended UE, the DCI format \texttt{1\_0} is cyclic redundancy check~(CRC) scrambled with the UE's random access radio network temporary identifier~(RA-RNTI), allowing the UE to recognize the message as relevant to its RACH procedure. If the UE receives a DCI format \texttt{1\_0} which has its CRC bits scrambled by the RA-RNTI, the UE proceeds to decode the transport block within the corresponding PDSCH resource allocation~\cite{5GNRRand98:online}. In the simplest scenario, where no RF frontends (power amplifiers and low noise amplifiers) are attached to both the UE and gNB running the open-source 5G protocol stack, the DCI format \texttt{1\_0} is easily detected and decoded. In what follows, we highlight this monitoring, detection, and decoding process of DCI format \texttt{1\_0} in the no-RF-frontend scenario. Based on the \texttt{pdcch-ConfigCommon} in the RRC message, the UE begins monitoring search spaces for the DCI to decode. Generally, the UE monitors the first OFDM symbol of each DL slot as well as the special slots for the DCI since it has no foreknowledge of where gNB will schedule the DCI. The UE successfully detects and decodes the DCI format \texttt{1\_0} with its preamble index value. 

 \begin{figure}[h!]
    \centering
    \includegraphics[width=1\columnwidth]{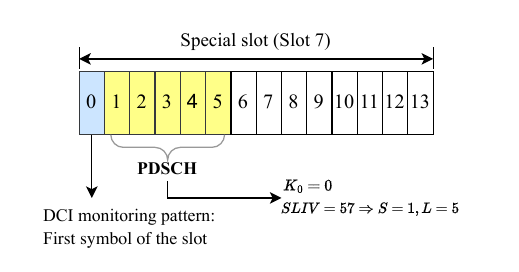}\vspace{-2mm}
    \caption{\texttt{msg2} Scheduling in a Special Slot}
    \label{fig:msg2_sched}
\end{figure}

After the DCI decoding process, the UE checks the \texttt{pdsch-TimeDomainAllocationList} provided in the \texttt{pdsch-config} in RRC to derive $k_0$ and the Start Symbol and Length Indicator Value~(SLIV) for the PDSCH resource containing the Random Access Response~(RAR). The parameter $k_0$ signifies the slot offset between the DL slot where PDCCH(DCI) for downlink scheduling is received and the DL slot where PDSCH data is scheduled~\cite{5GNRK0K171:online}. The SLIV represents the start symbol and the number of consecutive symbols or the length of the \texttt{msg2} PDSCH. The values $k_0$ and SLIV are shown in Fig.~\ref{fig:msg2_sched}. Algorithm~\ref{alg:sliv} describes the procedure for 
calculating SLIV from the start symbol index~($S$) and the length~($L$). The derivation of the $S$ and~$L$ from  SLIV is non-trivial. Therefore, it would be handy to use the lookup table presented in~\cite{SLIV_table:online}.
OAI 5G protocol stack uses an SLIV value of 57 for \texttt{msg2} scheduling, i.e., OAI uses 5~downlink OFDM symbols to schedule \texttt{msg2} PDSCH starting from the second DL symbol. The UE, after decoding the \texttt{msg2} PDSCH carrying Random Access Response~(RAR), prepares \texttt{msg3} to be transmitted to the gNB on PUSCH.

\begin{algorithm}[t!]
\caption{SLIV Calculation}
\label{alg:sliv}
\begin{algorithmic}[1]
\Statex \textbf{Inputs:}
\Statex \ $S$---Start Symbol Index 
\Statex \ $L$---Number of Consecutive Symbols 
\Statex \textbf{Output:} $SLIV$
\If {$(L-1) \leq 7$}
    \State $SLIV \gets 14 \times (L-1) + S$
\Else
    \State $SLIV \gets 14 \times (14 - L + 1) + (14 - 1 - S)$
\EndIf\\
\Return $SLIV$
\end{algorithmic}
\end{algorithm}

\subsection{Monitoring and Detection of DCI Format 1\_0 with RF Frontend}
\begin{figure}[h!]
    \centering
    \includegraphics[width=\columnwidth]{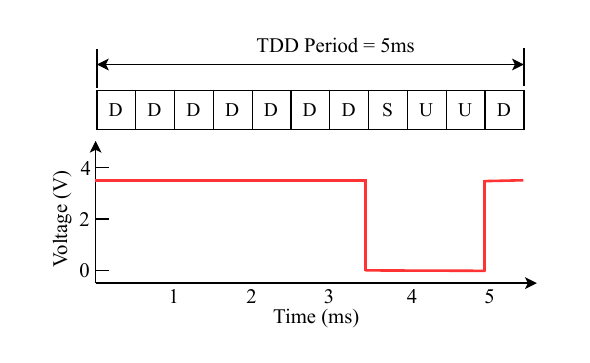}\vspace{-3mm}
    \caption{GPIO Control Signal Switching in OAI gNB}
    \label{fig:gpio_switch}
\end{figure}

Interfacing SDRs with RF frontends (i.e., PAs and LNAs) introduces challenges 
into the 5G RACH procedure, specifically the detection of DCI format \texttt{1\_0}. For instance, as explained in Section~\ref{sec:oai_signal_transmission}, the RF frontend 
is only in transmission mode (with PA activated) when it receives a 3.3V control signal from the SDR GPIO; otherwise, the RF frontend continues to be in reception mode (with LNA activated).
Given the OAI TDD configuration of \texttt{DDDDDDDSUU} specified in Section~\ref{subsec:tdd_ul_dl_config}, the OAI protocol stack uses the UHD \texttt{set\_gpio\_attr} API to set the RF frontend 
in transmission mode with 3.3V GPIO signal for 7 consecutive DL slots. Immediately after the last DL slot's last symbol, the GPIO control signal is set to 0V to switch the RF frontend 
to 
reception mode for the remaining three slots, as shown in Fig.~\ref{fig:gpio_switch}. 
That is, the special slot is treated as an UL slot in the OAI 5G protocol stack. Since the gNB schedules the DCI to be transmitted in the first DL symbol of the special slot and the \texttt{msg2} RAR in the remaining DL symbols of the special slot, the PA does not actually transmit the DCI and RAR on PDCCH and PDSCH, respectively, even though the software stack schedules them for transmission. This is because, for the entire duration of the special slot, the LNA is activated instead of PA being activated for the duration of the DL symbols and LNA for the remaining duration of the slot. 
As a result, \texttt{msg2} is not transmitted by the~PA, even though the OAI scheduler schedules it for transmission. Subsequently, the OAI UE never receives \texttt{msg2},  thereby terminating the random access procedure with a ``\textit{RAR reception failed}'' message. Continuous \texttt{msg2} reception failures form the basis of the 5G random access procedure problem in real-world  implementations of the open-source 5G stacks with USRPs 
at UEs and gNBs.

One \textit{naive} approach to solving the 5G random access problem is to modify the OAI code in a way that the GPIO control signal is in sync with the data transmission and reception. In other words, by precisely tuning the GPIO control signal such that within the special slot, the PA only transmits during the DL symbols, and the LNA is only activated during the UL symbols of the special slot. The downside of this approach is that only a few OFDM symbols are available to schedule \texttt{msg2} and \texttt{msg3}  (i.e., 6 and 4 OFDM symbols, respectively, by default). As we will show 
with experiments in Section \ref{subsec:sliv}, using fewer symbols to schedule \texttt{msg2} and \texttt{msg3} leads to a highly unreliable 5G random access procedure, especially in large cell scenarios.


\section{Leveraging Full DL and UL Slots for 5G Random Access}




To solve the 5G random access problem, we propose an approach that schedules \texttt{msg2} in a full DL slot, specifically the last DL slot in the TDD period. The full DL slot has all 14~OFDM symbols which gives the flexibility of scheduling \texttt{msg2} with more OFDM symbols than the special slot can provide. 
Scheduling \texttt{msg2} in a full DL slot requires modifying the way in which \texttt{msg2} is scheduled at the gNB using the \texttt{nr\_schedule\_\texttt{msg2}} function. Originally, OAI implemented this function to schedule \texttt{msg2} always in a special/mixed slot when there exists a special slot in the TDD slot pattern or configuration. We modify this scheduling strategy such that the gNB always schedules \texttt{msg2} in the last DL slot, irrespective of the TDD slot configuration. As per 3GPP specification 38.214~\cite{Specific64:online}, UE is scheduled to transmit \texttt{msg3} on PUSCH a number of slots after the last symbol of \texttt{msg2} is received. With reference to the number of slots for a PUSCH transmission scheduled by a RAR UL grant, if a UE receives a PDSCH with a RAR message ending in slot $n$ for a corresponding PRACH transmission from the UE, the UE transmits the PUSCH in slot $n_1$ given by $n_1 = n + k_2 + \Delta$, where $k_2$ is specified in the \texttt{push-TimeDomainAllocationList} sent to the UE in the SIB1 message~\cite{RandomAc37:online}. The value of $\Delta$ is chosen based on the numerology order provided in TABLE~\ref{tab: delta}.


 \begin{figure}[h!]
    \centering
    \includegraphics[width=1.0\columnwidth]{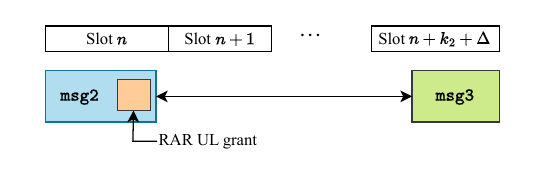}\vspace{-3mm}
    \caption{\texttt{msg3} Scheduling after \texttt{msg2} Reception}
    \label{fig:msg3_sched}
\end{figure}

\begin{table}[h!]
\centering
\caption{Definition of $\Delta$}
\renewcommand{\arraystretch}{1.2} 
\begin{adjustbox}{max width=1.0\columnwidth}
\begin{tabular}{| c | c | c | c | c |}
\hline
\textit{\(\mu_{PUSCH}\)}         & 0 & 1 & 2 & 3 \\\hline
\(\Delta\) & 2 & 3 & 4 & 6\\\hline
\end{tabular}
\end{adjustbox}
\label{tab: delta}
\end{table}

    \begin{figure*}[htbp!]
    \centering
    \includegraphics[width=\textwidth]{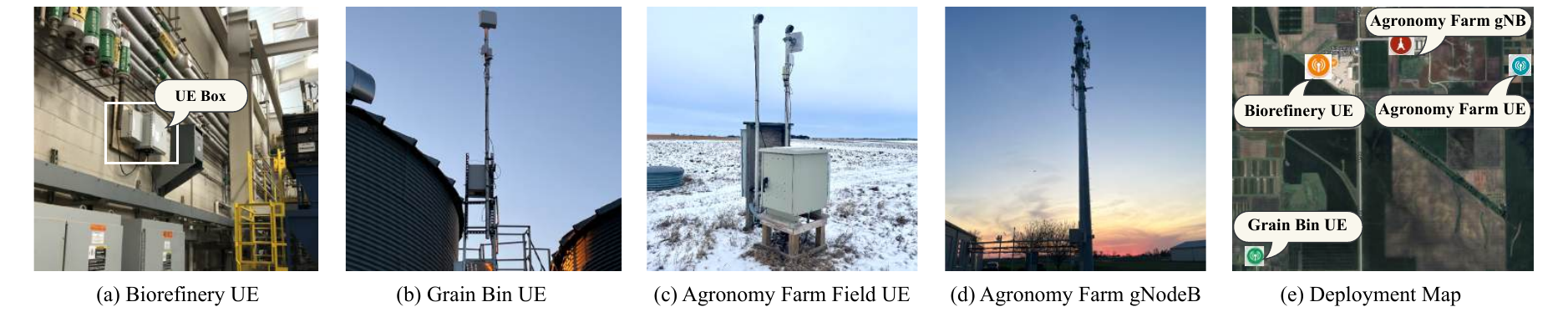}
    \caption{Agronomy Farm gNB and Surrounding UEs}
    \label{fig:deploy_figs}
\end{figure*}

 \begin{figure}[h!]
    \centering
    \includegraphics[width=1.0\columnwidth]{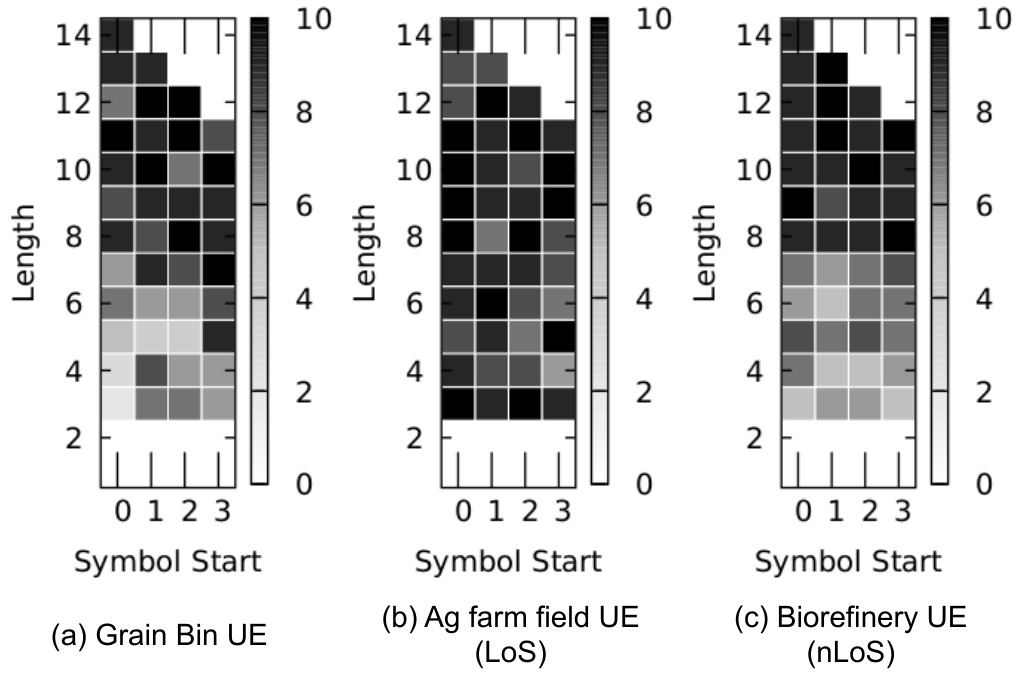}
    \caption{Effect of different symbol start and length combinations on \texttt{msg2} reception for different deployment scenarios}
    \label{fig:msg2_sliv_plot}
\end{figure}

By design, the OAI protocol stack uses $k_2 = 7$, i.e., the gNB schedules the UE to transmit \texttt{msg3} after 10 slots. If the same implementation is used after scheduling \texttt{msg2} in the last DL slot, which is the slot index~6 as per the TDD slot configuration \texttt{DDDDDDDSUU}, \texttt{msg3} will be scheduled in slot index 16, which is a DL slot. Therefore, we modify the RRC configuration to add a new $k_2$ entry to the \texttt{push-TimeDomainAllocationList} such that \texttt{msg3} will be transmitted in a UL slot instead. More specifically, we use $k_2 = 9$ to schedule \texttt{msg3} in the UL slot index~18. Given the large cell radius of the SDR deployment on ARA, there is a high probability for some degree of path loss and interference as signal quality degrades over longer distances. To ensure the reliability of \texttt{msg3} transmission and reception and minimize retransmissions, we schedule \texttt{msg3} with a relatively large number of symbols by default. To this end, we add a new entry of \texttt{startSymbolAndLength} in the \texttt{push-TimeDomainResourceAllocation} of the SIB1 message. In the next section, we evaluate the performance of using different start symbols and length values to schedule \texttt{msg2} and \texttt{msg3} in real-world settings.




\section{Evaluation of \texttt{msg2} and \texttt{msg3} Start Symbols and Length Values on 5G Random Access Procedure  }
\label{subsec:sliv}

Our approach of utilizing the full DL and UL slots for scheduling \texttt{msg2} and \texttt{msg3} makes the whole 14 OFDM symbols available within the slot. Therefore, we have the flexibility of choosing the OFDM start symbol index and the number of consecutive OFDM symbols to schedule \texttt{msg2} and \texttt{msg3}.
In this section, we evaluate the impact of different start symbols and lengths of \texttt{msg2} and \texttt{msg3} on the success of the 5G random access procedure. We evaluate the impact through real-world experiments leveraging the open-source 5G deployment of ARA wireless living lab~\cite{ARA:Design-Impl}. 

\subsection{Experimental Setup}
\label{sec:expt_setup}

The ARA wireless living lab features seven SDR gNBs, called \textit{AraSDR}~\cite{arasdr-2024},  running open-source 5G software stacks. AraSDR is deployed around the cities of Ames, Gilbert, Boone, and Nevada, covering an area of diameter 30\,km across central Iowa. 
In addition, 20~fixed-location UEs are deployed in rural settings, i.e., in crop/livestock farms, grain bins, city utility service buildings/vehicles, and small industrial setups. 
Each gNB is equipped with a compute server that hosts three SDRs and enables 5G experimentation through Docker containers. 

The software framework of AraSDR is 
based on OpenStack cloud operating system, and it allows users to reserve compute and wireless resources to execute experiments. The framework offers container-based resource provisioning where pre-built containers can be used for running open-source 5G gNB, UE, and the core network, thus enabling fully reproducible experiments. 
The 5G core network runs on the compute node in the data center.
Once the reservations for gNBs and UEs are made and containers are launched, both UEs and gNBs are configured to establish a wireless link between them using~OAI. The UE is registered with the core, and a tunnel is established via the wireless link from UE to the gNB and to the core network via virtual endpoints attached to the gNB container by the OpenStack \emph{Neutron} module (see \cite{arasdr-2024} for further details on the software framework).
Fig.~\ref{fig:deploy_figs} shows a part of the AraSDR deployment, i.e., Agronomy Farm gNB and surrounding UEs used in this study.
\begin{figure*}
\begin{minipage}[t]{0.3\textwidth}
  \includegraphics[width=\linewidth]{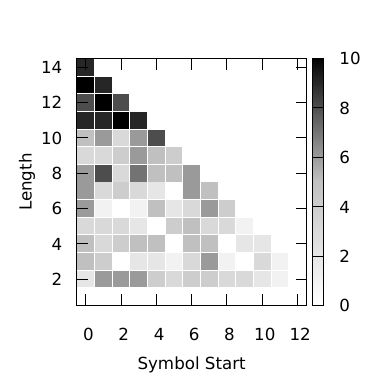}
  \caption{\texttt{msg3} reception probability for Grainbin UE}
  \label{fig:msg3_gbin}
\end{minipage}%
\hfill 
\begin{minipage}[t]{0.3\textwidth}
  \includegraphics[width=\linewidth]{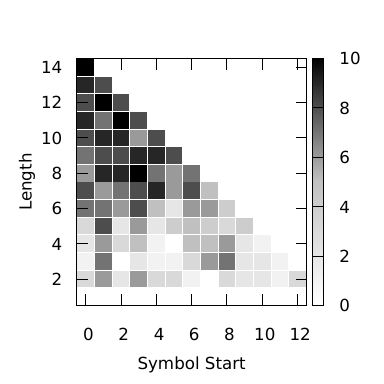}
  \caption{\texttt{msg3} reception probability for Ag farm UE (LoS from gNB)}
  \label{fig:msg3_ague2}
\end{minipage}%
\hfill
\begin{minipage}[t]{0.3\textwidth}
  \includegraphics[width=\linewidth]{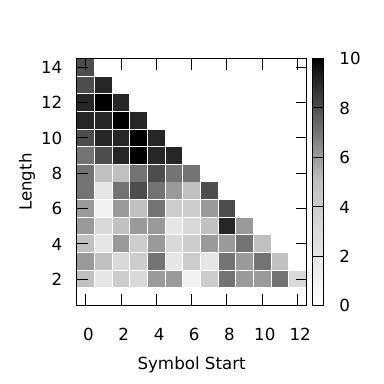}
  \caption{\texttt{msg3} reception probability for Biorefinery UE (nLoS from gNB)}
  \label{fig:msg3_BCRF}
\end{minipage}%
\vspace{-4mm}
\end{figure*}

\subsection{Experiment Scenarios}
We present a detailed analysis of how the time-varying wireless channel affects the detection of \texttt{msg2} and \texttt{msg3} scheduling on PDSCH and PUSCH, respectively, for different symbol start and lengths. 
To understand this phenomenon, we present two sets of experiments---Scenario~A and Scenario~B.

To foster the adoption of open-source 5G and NextG software stacks in commercial settings, it is essential to understand their performance across various deployment scenarios, including different distances and both line-of-sight~(LoS) and non-LoS~(nLoS) conditions between gNBs and UEs. Results presented in \cite{OAImeascampaign} and \cite{OAIperfindoor} showed that obstacles such as walls reduce the strength and quality of 5G signals, and subsequently lowering the data rate. As the first ever study to present the effect of distance and obstacles on the open-source 5G random access procedure, we utilize a single gNB at the Agronomy Farm and three UEs spatially distributed around the gNB. One UE is located at the Grain Bin, 1600\,m from the gNB; another UE is located at the Agronomy Farm field, 650\,m from the gNB; and the third UE is situated at ISU's Biorefinery facility, an industrial setting 450\,m from the gNB. 
In Scenario~A, we vary the symbol start and lengths for \texttt{msg2} and \texttt{msg3}, to analyze the effect of channel fading on the detection of \texttt{msg2} and transmission of \texttt{msg3} for UEs located at the Grain Bin and Agronomy Farm field, each at different distances from the gNB. In Scenario~B, we run similar experiments; however, instead of focusing on UEs at varying distances from the gNB, we examine UEs positioned in LoS and nLoS conditions relative to the gNB. Specifically, we use the UE at the Agronomy Farm field, which is in LoS with the gNB, and the UE located at the Biorefinery facility where there are several obstacles in between the UE and gNB representing the nLoS scenario.

We conduct experiments in both scenarios simultaneously using a single gNB  located at the Agronomy Farm. For every experiment under each scenario, we run 10~iterations with the same SLIV value (i.e., start symbol and length combination) and present the number of times we successfully receive \texttt{msg2} and \texttt{msg3} at the specific UE and gNB, respectively.

\subsection{Observations and Discussion}
\subsubsection{Scenario A}

Fig.~\ref{fig:msg2_sliv_plot}(a) shows the number of times \texttt{msg2} is detected at the Grain Bin UE for various SLIV combinations. Since a mapping \textit{typeA} is used for \texttt{msg2} scheduling on PDSCH, only symbol indices 0--3 can be used as symbol starts. However, the number of consecutive OFDM symbols that can be used to schedule \texttt{msg2} is given by $L = 14 - k$, where $k$ is the symbol start index. From Fig.~\ref{fig:msg2_sliv_plot}(a), it can be observed that the likelihood of \texttt{msg2} being detected by UE increases with relatively larger length values. This is because for UEs located farther from the gNB, the signal quality degrades significantly, and the SINR deteriorates, adversely affecting the symbol detection and decoding. Therefore, scheduling \texttt{msg2} with larger length values increases the correlation window for successful receiver detection in severe fading conditions. Fig.~\ref{fig:msg2_sliv_plot}(b) shows how often the UE at the Agronomy Farm field successfully detects \texttt{msg2} given the same wireless channel temporal conditions~(i.e., both experiments were run simultaneously). It can be observed that when the UE is located close to the gNB, the successful detection of \texttt{msg2} on PDSCH at the UE is not affected by the number of consecutive symbols used for \texttt{msg2} scheduling. For instance, \texttt{msg2} is often detected when scheduled with OFDM symbols ranging from 3 to 13.

Under the same conditions and using the same experimental setup, we collect data to analyze the count on the successful reception of \texttt{msg3} at the gNB on PUSCH. It is worth noting that the OAI software stack leverages PUSCH mapping \textit{typeB}. Also, given the condition of modulation and coding scheme~(MCS) index not exceeding 28 and the minimum size of the \texttt{msg3} transport block not less than 7~bytes, the symbol length for \texttt{msg3} scheduling must be no less than 2~symbols. In this experiment, we consider a successful reception of \texttt{msg3} as a first-time reception or any reception within the \texttt{msg3} retransmission window, which is three consecutive frames, each with a duration of 10\,ms. Fig.~\ref{fig:msg3_gbin}, shows the number of times \texttt{msg3} is successfully received by the gNB from a UE located 1600\,m away for all SLIV combinations. It can be seen that using a larger number of consecutive symbols or length value, i.e., 11~symbols and above, leads to a higher chance of successful \texttt{msg3} reception. The reason for this observation is the same as that of the \texttt{msg2} detection case---smaller length values are more affected by signal degradation caused by the wireless channel over longer distances. For the nearby UE scenario in Fig.~\ref{fig:msg3_ague2}, using smaller length values below 8~symbols leads to a very low and near-zero probability of successfully detecting \texttt{msg3}. This observation is due to the fact that when using open-source 5G software stacks with SDRs, the gNB receive chain is highly sensitive to interference and can also be easily saturated with high signal strengths, i.e.,  \texttt{msg3} with smaller length values is more prone to corruption compared to those with larger lengths.

\subsubsection{Scenario B}
It is evident from Fig.~\ref{fig:msg2_sliv_plot}(b) and Fig.~\ref{fig:msg2_sliv_plot}(c) 
that 
the UEs situated at the Agronomy Farm field and the Biorefinery facility 
exhibit different probabilities of successfully receiving the \texttt{msg2} for various symbol start and length combinations, despite being at similar distances from the gNB.
For instance, the UE at the Agronomy Farm field has 80\%--100\% chance of successfully detecting \texttt{msg2} when it is transmitted with at least three consecutive symbols. However, in the nLoS setting depicted in Fig.~\ref{fig:msg2_sliv_plot}(c),
to maintain the same success probability
(i.e., at least 90\%), 
\texttt{msg2} must be transmitted with at least eight consecutive OFDM symbols. This is 
due to the obstructions at the Biorefinery facility, which reduces the quality and strength of the 5G signal. Therefore, it is important to use more symbols to transmit \texttt{msg2} for achieving a higher success rate of detection in nLOS settings. 

Similar insights can be drawn from Figs. \ref{fig:msg3_ague2} and~\ref{fig:msg3_BCRF} where relatively smaller length values improve the \texttt{msg3} reception probability 
for the UE at the Agronomy Farm field, which is in LoS with the gNB compared to that at the Biorefinery facility. It can be inferred that to achieve a higher \texttt{msg3} reception rate~(90\%--100\%) in a significantly obstructed environment,
at least nine consecutive OFDM symbols must be used to transmit \texttt{msg3}. As mentioned in Scenario~A, for both UEs, using smaller length values or fewer OFDM symbols to transmit \texttt{msg3} results in a slight chance of successful reception at the gNB, which is due to the fact that the SDR receive chain is susceptible to high signal levels and surrounding noise when the UEs are close to the gNB. Consequently, \texttt{msg3} becomes corrupted when transmitted with too few OFDM symbols.

\section{Concluding Remarks} 

In this paper, we presented the problem of random access procedure failure on outdoor programmable wireless living labs using OpenAirInterface5G software stack and software-defined radios. We highlighted the general implementation of 5G random access procedure in open-source 5G software stacks. We also presented an overview of how special slots are utilized in the open-source 5G random access procedure and the corresponding performance issues observed when interfacing SDRs with TDD amplifiers. Moreover, we presented an approach to solve the problem presented using full uplink and downlink slots for the 5G random access procedure. Finally, we leveraged outdoor 5G experiments using OAI on the ARA wireless living lab to understand and analyze the effect of the dynamic wireless channel on different start and symbol length combinations used in scheduling \texttt{msg2} and \texttt{msg3} on PDSCH and PUSCH respectively. 
The results presented provide insights into how we can better optimize open-source 5G random access procedure in different real-world deployment scenarios such as those highlighted in this paper. This is essential in the design and prototyping of open-source 5G and next-generation cellular networks and applications that are at par with commercial counterparts in terms of reliability, coverage, and resource efficiency.


\section*{Acknowledgement}
This work is supported in part by the NIFA award 2021-67021-33775, and NSF awards 2130889, 2112606, 2212573, 2229654, 2232461. We thank Francesco Mani, Cedric Roux, and other colleagues at Eurecom for their help in understanding certain implementation choices of the OAI 5G random access procedure.

\balance

\end{document}